\documentstyle[11pt,newpasp,twoside,epsf]{article}
\def\edcomment#1{\iffalse\marginpar{\raggedright\sl#1\/}\else\relax\fi}
\reversemarginpar

\begin{document}
\title{AGN Unification: An Update}
\author{C. Megan Urry}
\affil{Yale University, Yale Center for Astronomy \& Astrophysics,
P.O. Box 208121, New Haven, CT 06520-8121}

\begin{abstract}
The paradigm for AGN unification is reviewed, in terms of its
optical manifestation as obscuration in the equatorial plane
and its radio manifestation as relativistic beaming of the 
jet emission. Within this paradigm, observed AGN properties 
depend strongly on orientation angle. The predictions of unification
are commensurate with the local numbers of AGN of various types. 
The outstanding question concerns a possible population of
obscured AGN at high redshift, which are thought to produce
the X-ray background, although few are observed directly. 
We describe early results from the 
{\em HST}-{\em SIRTF}-{\em Chandra} GOODS survey,
which has as one of its principal goals the search for such obscured AGN.
\end{abstract}

\vspace{-0.3cm}

\section{Introduction}
\label{sec:intro}

Based on nearly 20 years of detailed investigations of active galaxies, 
a unification paradigm has emerged in which active galactic nuclei share
certain fundamental ingredients (illustrated in Fig.~1):\\
$\bullet$~a {\bf supermassive black hole}, $10^{6-10} M_\odot$;\\
$\bullet$~an {\bf accretion disk and corona}, heated by magnetic and/or
viscous processes so that it radiates at optical through soft X-ray energies;\\
$\bullet$~high velocity gas, usually referred to as 
the {\bf broad-line region};\\
$\bullet$~lower velocity gas in the {\bf narrow line region};\\
$\bullet$~an {\bf obscuring torus} (or other geometrical form) 
of gas and dust, hiding the broad-line region from some directions; and\\
$\bullet$~a {\bf relativistic jet}, formed within $\sim100$ Schwarzschild radii
of the black hole, and extending outward for tens of 
kpc, and in some cases, as much as a Mpc.\vspace{0.1cm}\\
Except perhaps for the last item, the unification paradigm holds 
that all AGN have the same ingredients, though certainly with 
intrinsic variations in black hole mass, ionization 
parameter, size, density, luminosity, etc.
According to unification, many of the principal observational 
characteristics of AGN --- such as overall spectral energy distribution 
and emission line type (broad- or narrow-line) --- stem from 
orientation rather than some fundamental intrinsic difference.

\begin{figure}
\label{fig:schematic}
\plotfiddle{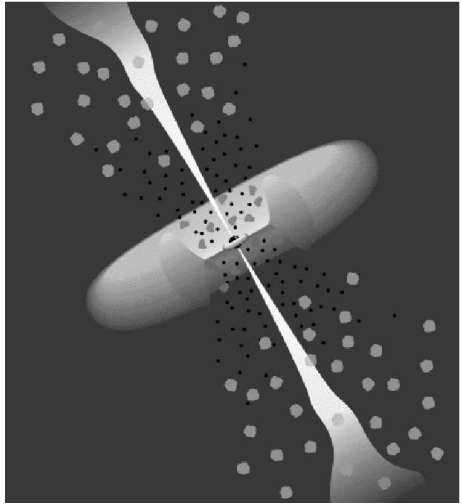}{2.5in}{0}{100}{100}{-100}{-20}
\caption{Schematic picture of a radio-loud AGN (not to scale),
with collimated radio jets escaping along the axis of a thick torus of dust
and gas. Depending on orientation, the torus hides or reveals the luminous
continuum emission from the central supermassive black hole and its
surrounding accretion disk. Relativistic beaming of the jet radiation
generates an aspect-dependent appearance along the same axis. 
\vspace{-0.1cm}
}
\end{figure}

Relativistic jets are almost certainly pre\-sent in all radio-loud 
AGN\footnote{Radio-loud AGN are those with a high ratio of radio to 
op\-ti\-cal emis\-sion, com\-mon\-ly de\-fined by the ratio between
5\,GHz and 5000\,\AA\ flux, $\log(F_{\rm 5\,GHz}/F_{\rm 5000\,\AA})>1$}
though with a range 
of intrinsic kinetic powers. The weaker jets decelerate relatively 
close to the central engine, often within the galaxy, while more 
powerful jets plow through the interstellar medium and into the 
intergalactic medium before forming large-scale radio lobes. 
Much ink has been devoted to the question of why some AGN are 
radio-quiet and others radio-loud (roughly 5-10\% are radio-loud, 
although the fraction increases with AGN and/or host galaxy luminosity). 
These can be divided into ``apples and oranges" theories --- 
i.e., that radio-loud and radio-quiet AGN are actually different beasts, 
with some defining difference such as black hole mass, or magnetic 
field near the black hole --- and ``Macintosh and Macoun" theories 
--- i.e., that fundamentally the AGN are similar, but the combination of jet power 
(which has an intrinsic spread) and host galaxy density (ditto) 
conspires to permit the formation of a bright radio source or to 
squelch the jet before it propagates far enough to be visible; see 
\S~5. 

Obscuration is less controversial but more poorly known. 
Local AGN obscuration has been well studied, starting with 
the spectropolarimetric observations of Antonucci \& Miller 
(1985), which suggested the obscuration occupies a 
substantial solid angle and blocks the broad-line region 
and accretion disk from certain lines of sight. 
That is, the distinction between type~1 AGN (which have both broad and 
narrow emission lines) and type~2 AGN (only narrow emission 
lines) appears to be largely one of orientation 
with respect to the line of sight, at least locally. 
(There is still some controversy about this simple picture; 
in particular, whether torus geometry depends on luminosity 
and whether excess starburst activity is associated 
with type~2 AGN; see Veilleux 2003.) 

At higher redshift, the situation is much less clear. 
A few luminous type~2 AGN have been found at $z\sim2$--3
(e.g., Stern et al.\ 2002; Norman et al.\ 2002), 
but certainly not as many as type~1 quasars found 
in optical/UV surveys at these redshifts. 
One of the outstanding questions in AGN research today 
is the question of obscured AGN at the ``quasar epoch" --- 
how common are they relative to type~1 AGN?
The next few years will certainly provide an answer, with
deep multiwavelength surveys like GOODS, SWIRE, and others (\S~4).

\vspace{-0.3cm}

\section{The Unification Paradigm for Active Galaxies}
\label{sec:paradigm}
\vspace{-0.1cm}

Active Galactic Nuclei go by a long list of names:
AGN, quasars, QSO (quasi-stellar object), blazars, BL Lac objects, 
Seyfert galaxies, radio galaxies, FR1 (Fanaroff-Riley type~1), FR2, etc.
The unification paradigm,
as described above, holds that 
orientation-dependent observational differences define these 
categories, rather than intrinsic differences.
Originally, 
``quasar'' meant radio loud (from quasi-stellar radio source) 
and ``QSO'' meant radio quiet; subsequently
the term quasar has been used to refer to all luminous AGN.
Some define an operational distinction
between ``AGN" and ``QSO" or ``quasar" 
(e.g., $M_V < -23$~mag means a quasar); 
however, AGN properties are similar and continuous from 
low to high luminosities, making this cut completely arbitrary.
Furthermore, this distinction is usually made in optical luminosity, 
which constitutes a highly variable fraction of the bolometric 
luminosity ($\sim$10\% in type~1 AGN, $\ll10$\% in type~2 AGN).
This makes the AGN/quasar distinction artificial,
so we use the term AGN for all luminosities.

It is important to note that there are really two 
separate unification schemes by orientation: optical and radio. 
The optical scheme explains the presence or absence of broad
line emission by the orientation of the obscuring torus, 
while the radio scheme explains the core-dominated (flat-spectrum) 
versus lobe-dominated (steep-spectrum) radio-loud AGN by orientation
with respect to the jet axis. 
The two schemes are related in that the axis of symmetry is almost certainly 
the same, but the viewing angle dependences are quite different, 
one dominated by optical depth and structure of the 
obscuring dust and gas,
the other by the beaming pattern, which depends on the Doppler factor(s). 

The first scheme arose from spectropolarimetric observations that
showed broad lines in polarized light from Seyfert 2 galaxies,
as if the broad emission lines were being scattered into 
the line of sight (Antonucci \& Miller 1985). 
Further evidence that obscuration is likely 
to be important in AGN, at least out to redshift $z\sim1$ or so,
comes from the spectrum of the X-ray ``background,"\footnote{The 
term ``X-ray background" refers to a luminous all-sky X-ray glow 
that was unresolved by early 
satellite experiments. 
We continue to use this term even though deep Chandra imaging shows the emission
is the sum of contributions from extragalactic point sources, mainly AGN. 
}
which is very hard (Fig.~2), 
rising sharply in the 2--10~keV band and peaking (in $EF_E$) near 40~keV. 
Type~1 AGN have much softer X-ray spectra, and thus a 
simple summation of unobscured AGN spectra would not produce the observed emission. 
Instead, it has been speculated that photoelectric absorption 
in the soft X-ray band hardens the X-ray spectrum of the sources 
that constitute the bulk of the X-ray background, 
implying there must be large numbers of obscured AGN 
at least out to $z\sim1$. (At higher redshifts, 
the observed radiation is rest-frame harder X-rays, increasingly unlikely 
to be affected 
by absorption.) Indeed, spectral synthesis 
models predict large numbers of obscured AGN at high redshift
(e.g., Gilli, Risaliti \& Salvati 1999; Gilli, Salvati \& Hasinger 2001), 
roughly 4--10 times as many as unobscured AGN, 
although these models do not match the observed 
redshift distribution of AGN (Hasinger 2002; cf., \S~4). 

\begin{figure}
\label{fig:xrbg}
\plottwo{XRBG_spec.epsi}{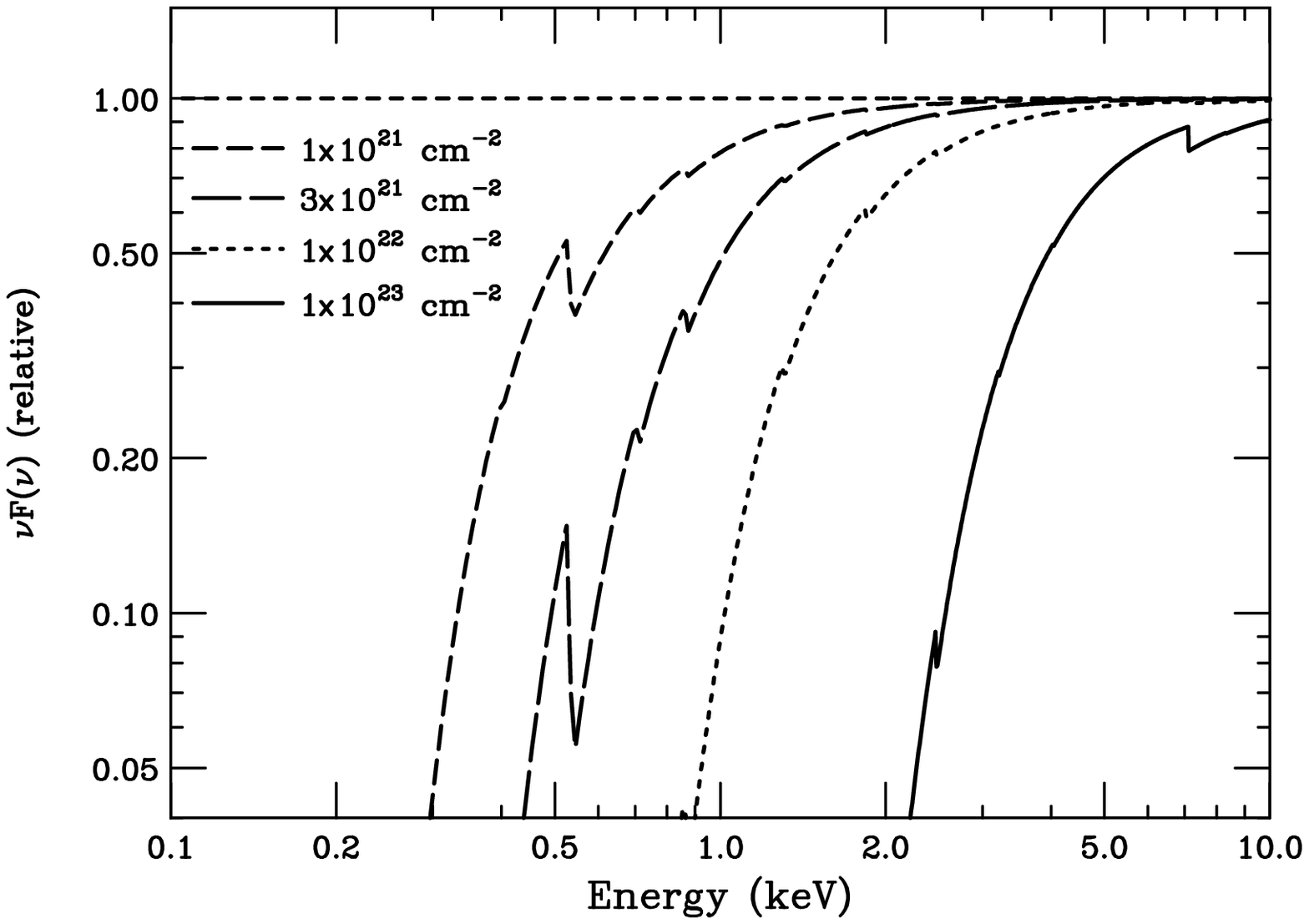}
\caption{The broad-band X-ray spectrum of the X-ray background
(thought to be the summed emission from many unresolved AGN) peaks
at 30--40~keV ({\it left panel}) --- much harder than type~1 AGN spectra,
which have approximately flat spectra in these units ({\it right panel, dotted line}).
Photoelectric absorption reduces the observed low-energy emission
from AGN ({\it right panel, curved lines}),
The summed X-ray emission from obscured AGN, with 
$N_H \geq 10^{22}$~cm$^{-2}$, can reproduce the spectrum of the
observed X-ray background if there are substantial numbers of
such AGN at $z\sim 1-3$ (Gilli et al.\ 1999, 2001).
}
\end{figure}

In some ways, the radio unification scheme is on firmer ground
(Urry \& Padovani 1995). 
There is abundant evidence for relativistic motion in the 
cores of flat-spectrum radio sources (the ones pointing toward us), 
as evidenced by:\\
$\bullet$~High brightness temperatures, 
in some cases well above the limit from the Compton catastrophe.\\
$\bullet$~Superluminal motion on parsec scales (Vermeulen \& Cohen 1994; 
Tingay et al.\ 1998) and, in M87, on the kpc scale as well 
(Biretta et al.\ 1995).\\
$\bullet$~The predominance of one-sided jets, 
as if Doppler beaming 
were acting even out to hundreds of kiloparsecs (Bridle \& Perley 1984).\\
$\bullet$~Depolarization asymmetries, i.e., the highly significant 
tendency for the greater Faraday depolarization of the large-scale 
radio lobes to occur on the counter-jet side, as if indeed the 
more distant lobe is on the side where the jet is moving away 
(Laing 1988; Garrington et al.\ 1988). 

The newest evidence in favor of relativistic beaming even on 
large scales is the discovery of X-ray emission from jets
on kiloparsec scales (Fig.~3).
In a recent Chandra survey of radio jets, 
Sambruna et al. (2002, 2003) showed that optical and X-ray emission 
from radio jets is common, and that for the higher redshift cases, 
it is well explained by a model in which relativistic electrons in the jet 
inverse-Compton scatter infrared photons from the cosmic microwave background (CMB). 
In the absence of relativistic beaming, the energy density 
of the CMB would be too low to be significant, but with bulk 
relativistic motion, the electrons ``see" a more intense photon field. 
In addition, after scattering, the high-energy radiation is also beamed. 
Thus for blazar jets (at small angles to the line of sight), 
particularly luminous sources at moderate to high redshift,
production of X-rays via this ``external Compton/CMB" model is quite plausible,
provided the bulk relativistic motion persists on large scales (tens to 
hundreds of kiloparsecs).
The optical radiation in these cases is usually interpreted 
at the high-frequency extension of the synchrotron spectrum.

\begin{figure}
\label{fig:jet}
\plotfiddle{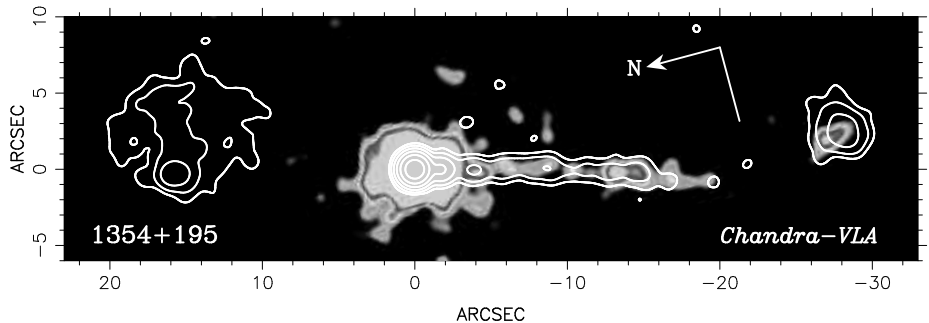}{1.35in}{-90}{50}{50}{-195}{215}
\caption{Example of a newly discovered kiloparsec-scale X-ray jet ({\it gray scale}),
with radio contours superimposed, from 
Sambruna et al.\ (2003). 
The X-rays follow closely the knots seen in the
radio image, and a hot spot in the jet-side radio lobe is also 
detected in X-rays. 
The likely explanation for the X-ray emission is inverse Compton
scattering of cosmic microwave background photons by the same 
relativistic electrons in the jet 
that emit synchrotron radiation in the radio. 
}
\end{figure}

Some of the jets discovered by Sambruna et al.\ are quite long 
($\sim100$~kpc), meaning de-projected lengths 
ten times larger for typical beaming parameters and orientation angles. 
The kinetic power in such jets is considerable. 

\section{Does AGN Unification Fit the Number Densities?}
\label{sec:numbers}

Because the Doppler beaming pattern is well known in the 
simplest uni-di\-rec\-tion\-al mo\-dels, it is possible to calculate 
the effect of beaming on the luminosity function 
(Urry \& Shafer 1984; Padovani \& Urry 1991). 
The effect is to distribute objects of the same intrinsic 
luminosity across a wide range of observed luminosities, 
flattening the slope of the luminosity function. 
The effect is greatest at low luminosities, where 
the numbers of objects are large, and then steepens 
to parallel the intrinsic luminosity function 
(at much higher observed luminosity).

Assuming the simplest plausible picture --- that each 
radio-loud AGN has a relativistic jet characterized 
by a single Doppler factor (one direction, one velocity), 
that each has a fixed fraction of total luminosity in 
intrinsic jet luminosity, and with a plausible distribution 
of Lorentz factors among jets 
--- it is straightforward 
to calculate the luminosity function that would be observed. 
In particular, we assume radio galaxies are roughly 
in the plane of the sky and thus not strongly affected 
by beaming, so that their luminosity function is essentially 
the parent luminosity function (with a small, calculable
correction for overall normalization). 
We then divide radio-loud AGN into steep-spectrum and 
flat-spectrum quasars, and predict the expected luminosity 
functions for each class, subject to known constraints on 
the ratio of core to extended radio flux (Padovani \& Urry 1992). 
Mean Lorentz factors $\gamma\sim10$,
compatible with the observed superluminal velocities,
fit the number densities very well (see Fig.~4). 

\begin{figure}[t]
\label{fig:lf}
\plotfiddle{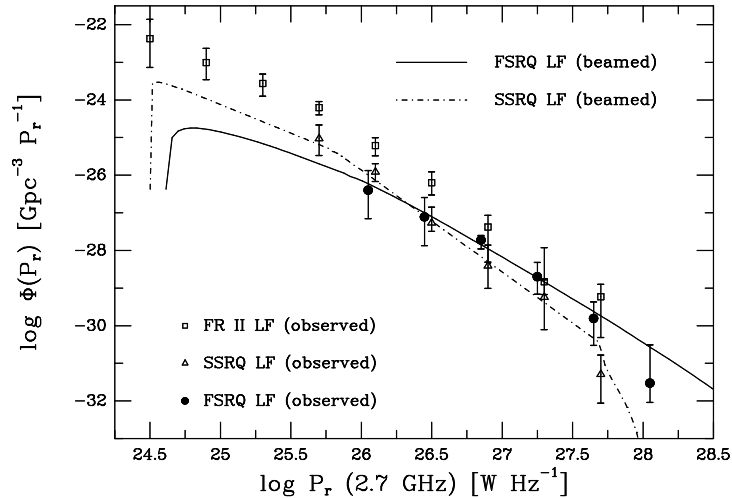}{2.45in}{-90}{40}{40}{-170}{210}
\caption{
Local differential radio luminosity functions for
flat spectrum radio quasars ({\it filled circles}),
steep spectrum radio quasars ({\it open triangles}), and FR~II galaxies
({\it open squares}).
The luminosity functions of FSRQ and SSRQ agree well
with the predictions of a beaming model calculated for an average 
bulk Lorentz factor $\sim10$; the range of angles to the line of sight is then
$\sim 0-14^\circ$ for FSRQ ({\it solid line}) 
and 14-38$^\circ$ for SSRQ ({\it dot-dashed line};
Urry \& Padovani 1992).
The similar slopes of the FR~II and SSRQ luminosity functions, and
the much flatter slope of the FSRQ luminosity function, are explained
very well by beaming. 
}
\end{figure}

\section{Obscured AGN at Redshift $>1$}
\label{sec:obsc}

Tests of optical obscuration unification are imminent 
now that surveys sensitive to obscured AGN are being carried out. 
The common method for finding quasars --- UV excess surveys --- 
is obviously a poor way to find UV-deficient AGN. 
Even the Sloan Digital Sky Survey (which can use unbiased 
color selection or spectroscopic selection) will mis-classify 
some obscured AGN as galaxies, since their optical nuclei 
will be completely obscured. 
(Surveys for narrow emission line objects are an exception, 
and have turned up some interesting results in the local universe; 
see Hao \& Strauss 2004, this volume.) 
In hard X-rays, however, AGN are the dominant population, and
the infrared should also be strong because of re-radiation 
of absorbed UV/soft X-radiation by the heated dust. 
Hence the issue of obscured AGN at moderate to high redshift 
is an ideal problem for the {\em HST} Treasury and {\em SIRTF} Legacy projects. 

Enter GOODS, the Great Observatories Origins Deep Survey, 
which was awarded roughly 800 hours of {\em SIRTF} observing time
and 412 orbits of {\em HST} time.
Because of the area and sensitivity of the Advanced Camera for Surveys, 
GOODS achieves nearly the depth of the Hubble Deep Field 
(Williams et al.\ 1996) but with roughly 100 times the area. 
The GOODS project targets the two deepest {\em Chandra} fields,
largely on the basis of the AGN investigation.
The {\em HST} data are now completed and mosaics are being made public; 
{\em SIRTF} was launched in August 2003;
GOODS {\em SIRTF} observations begin early in 2004. 

Early GOODS AGN papers include a discussion of z-band dropouts,
which might be highly obscured and/or high-redshift AGN (Koekemoer et al.\ 2003),
and an estimate of the space density of high-redshift AGN 
(Cristiani et al.\ 2003), essentially confirming earlier work on the 
Chandra Deep Field North (Barger et al.\ 2001). 
Our group has looked at the z-band magnitude distribution 
of X-ray sources (Fig.~5), 
which shows two peaks, brighter and fainter than $z\sim23$~mag. 
Obscured AGN can account for the faint peak providing one
recognizes that they are preferentially excluded from 
spectroscopic samples, and thus that observed redshift distributions
omit most of these (Treister et al.\ 2003).

\begin{figure}[t]
\label{fig:ACScounts}
\plottwo{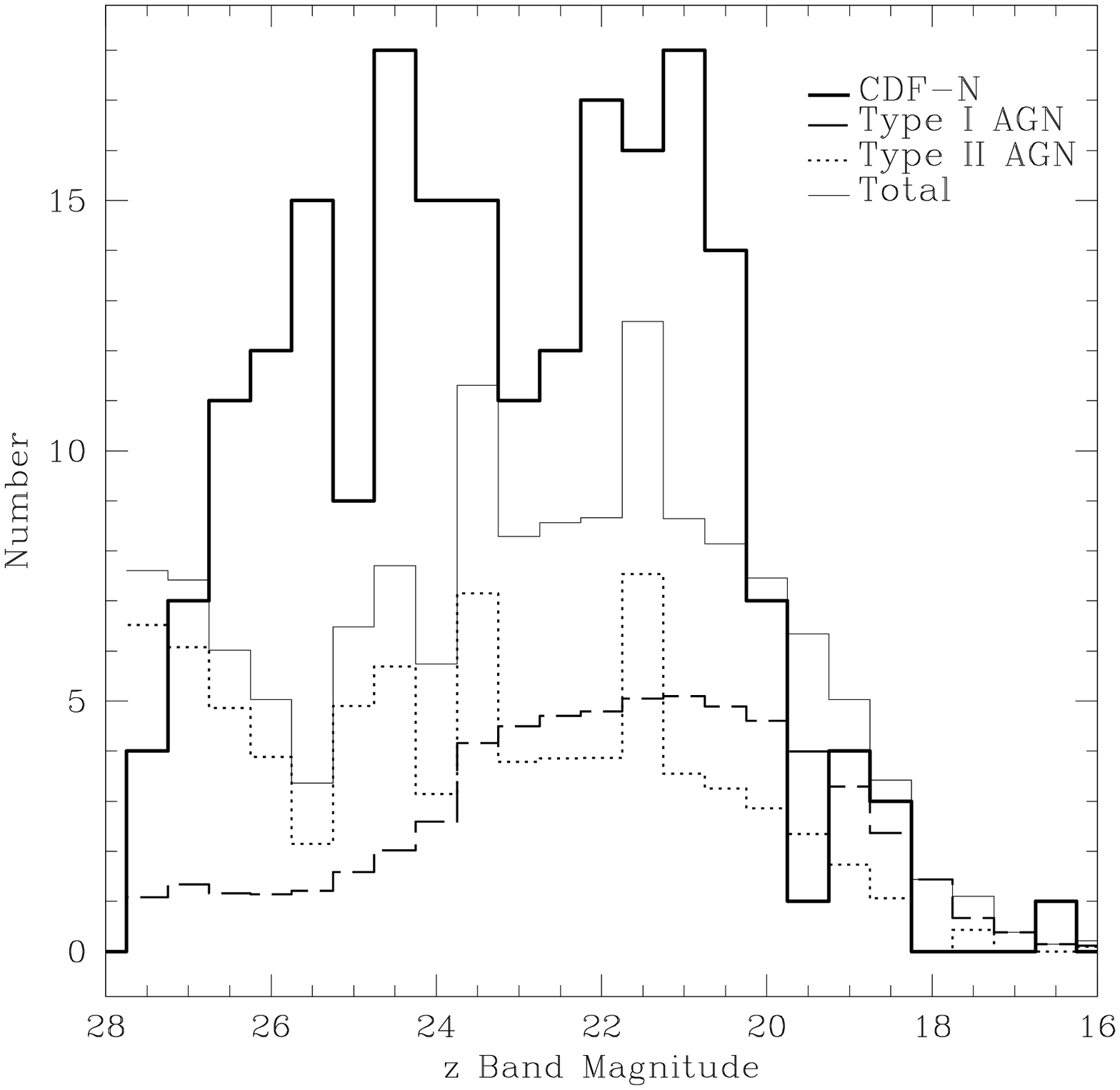}{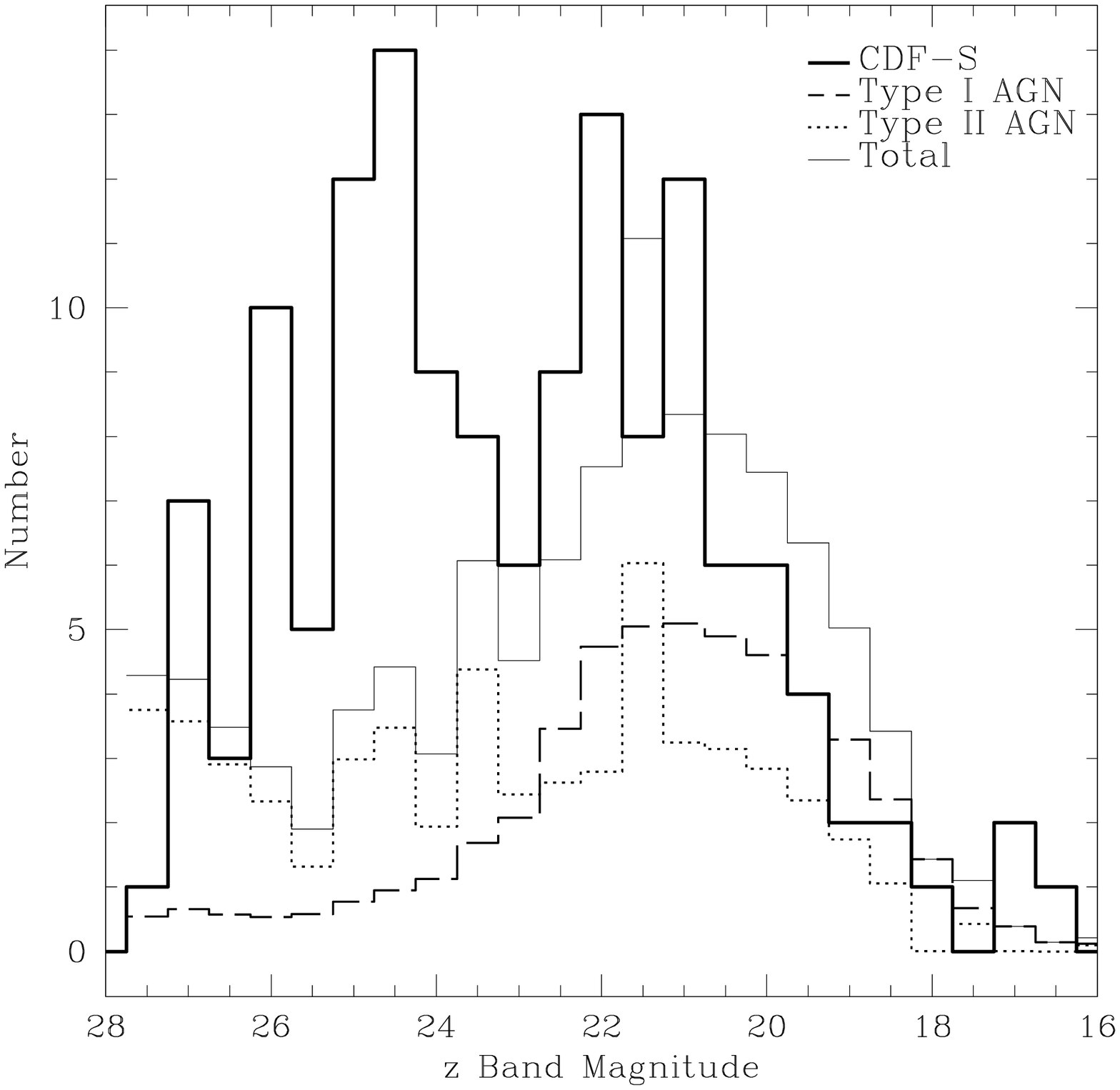}
\caption{Distribution of observed $z$-band magnitudes
for GOODS X-ray sources ({\it heavy solid lines}),
compared to the summed distribution ({\it light solid lines})
of Type~1 ({\it dashed lines}) and Type~2 ({\it dotted lines}) AGN
calculated from a hard X-ray luminosity function (Ueda et al.\ 2003).
This LF fails to account for the faintest sources ($z>23$) because of the
magnitude cut implicit in spectroscopic samples. A simple unification
model with many obscured AGN at all redshifts agrees well with the
observed distribution (Treister et al.\ 2003).}
\end{figure}

With its unique combination of ultra-deep X-ray ({\em Chandra}) and 
infrared ({\em SIRTF}) imaging, GOODS is ideal for finding obscured AGN, 
and the deep multicolor {\em HST} imaging allows
to separate the host galaxy and nucleus.
However, the GOODS area is small, and thus the volume too small 
to find typical luminous AGN, $L_X > 10^{45}$~erg/s.
To study luminous obscured AGN, larger survey areas are needed. 
This is one of the motivations for MUSYC, the MUltiwavelength 
Survey of Yale and Calan, a deep optical and infrared survey 
covering one square degree in four southern and equatorial fields. 
{\em Chandra}, {\em XMM}, and {\em SIRTF} imaging is or will be available for 
at least two of the four fields. The X-rays can be used to 
pinpoint likely AGN, and the infrared can be used to discriminate 
starbursts from AGN and to find partially obscured nuclei. 
The COSMOS 2-degree survey, with its X-ray, infrared, 
and optical coverage, will also find obscured AGN and enable detailed
study of their host galaxies as a function of AGN luminosity and
local galaxy density.
The SWIRE and MIPS GTO surveys with {\em SIRTF}, 
and the GEMS survey with {\em HST}, will also provide important
complementary coverage at different flux levels 
in the number counts, and thus should provide strong constraints 
on the cosmic evolution of the obscuration in AGN. 
Obviously SDSS is already making important contributions 
(Zakamska et al.\ 2003), as is the 2dF survey. 
Purely optical surveys, of course, will always 
be biased against obscured AGN.

These deep, large-area multiwavelength surveys address several 
questions: 
$\bullet$~{\bf AGN geometry.} What is the distribution and scale of the 
obscuring dust? Long time scale monitoring of infrared emission
(reverberation mapping of the dust) suggests the
physical scales are light-months (Clavel et al.\ 
1989), and ISO observations offer some constraints on the aspect ratio
(Clavel et al.\ 2000).\\
$\bullet$~{\bf Dependence of geometry on redshift or luminosity.}
There have been many suggestions that more luminous AGN have 
less obscuration, starting with Lawrence (1987),
in part because no luminous obscured AGN had been found at high redshift. 
Yet some of these arguments are based on surveys that would not have 
found obscured AGN in any case, given their blue bias.\\
$\bullet$~{\bf Demographics of black holes.} How many AGN are there at 
the ``quasar epoch"? AGNs are markers 
of black holes of substantial mass, and certainly of the 
peak accretion period. Possibly the period of maximum AGN 
activity represents the simultaneous growth of the black hole 
and the formation of the galaxy (Kormendy \& Gebhardt 2001).

One critical issue is estimating bolometric luminosities of AGN
when we know their emission is highly anisotropic. The observed
spectral energy distributions (SEDs) of neither type~1 nor type~2 
AGN are bolometric, at least if the unification paradigm is even
approximately correct. Instead, far-infrared and hard X-ray emission
are approximately isotropic (and thus a faithful representation
of the intrinsic emission) but the type~1s are over-luminous 
in the UV through soft X-ray, and type~2s are underluminous,
compared to the angle-averaged emission.
Given this, properly matching samples of 
type~1 and type~2 AGN (to test unification, for example)
is not as straightforward as many authors have assumed.
A hard X-ray-selected sample of AGN would avoid much bias, but
to date we do not yet have large areas done very deep with a 
high fraction of source identifications.
With Chandra, larger area surveys are now starting 
to be done, with perhaps the most obvious being the 
extended CDF-S area (GEMS).

\section{Host Galaxies and Black Hole Properties}
\label{sec:hg}

The ultimate aim in AGN studies is to understand 
the intrinsic physical parameters of AGN, starting with mass, 
accretion rate, and black hole spin --- quantities that may be 
obscured (pardon the pun) by orientation effects. 
Here we discuss AGN host galaxies and
review the current knowledge of black hole masses in
AGN, and how they relate (or do not relate) to observed AGN properties.

AGN host galaxies appear, in every way we can easily measure, 
to be like normal galaxies. For example, the Kormendy relation 
--- a projection of the fundamental plane onto the two morphological 
parameters, $r_e$ (half-light radius) and $\mu_e$ (surface 
brightness at the half-light radius) --- is the same for AGN 
as it is for normal galaxies (Fig.~6; 
Urry et al.\ 2000, 
O'Dowd et al.\ 2002, Taylor et al.\ 1996, Dunlop et al.\ 2003). 
Despite some claims to the contrary, it has been obvious 
for a while that AGN luminosity is largely de-coupled from 
host galaxy properties. Over a range of 5 orders of magnitude 
in nuclear luminosity, the host galaxies of radio-loud AGN 
(quasars and BL Lacs) cluster within one magnitude of the average value. 
There is no believable correlation between host galaxy luminosity 
and AGN luminosity (O'Dowd et al.\ 2002). 

\begin{figure}[t]
\label{fig:muere}
\plottwo{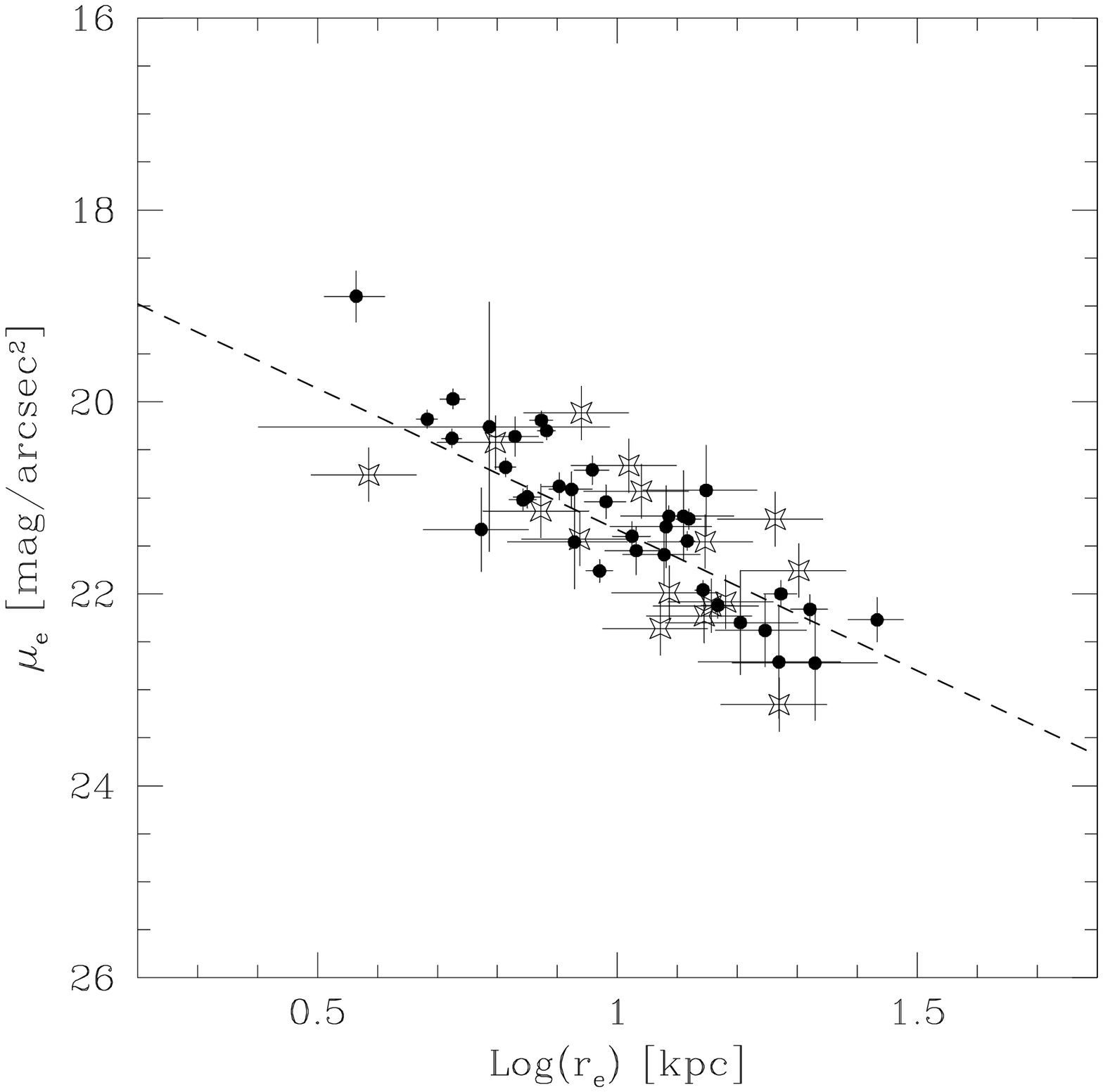}{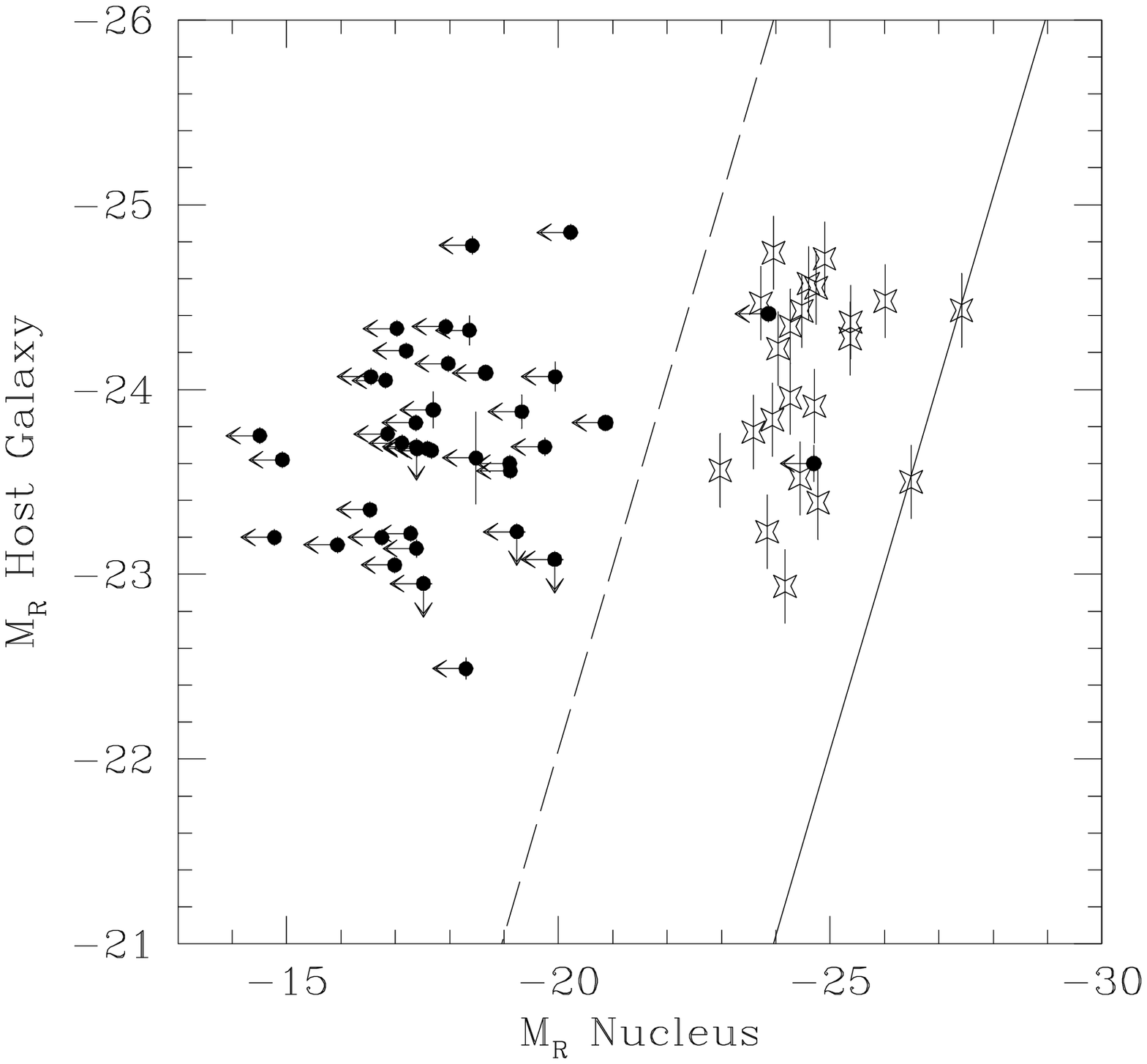}
\caption{{\it Left:} Surface brightness versus effective radius for 
the host
galaxies of radio-loud AGN ({\it symbols}) is well fit by Kormendy relation
for normal elliptical galaxies ({\it dashed line}; Hamabe \& Kormendy 1987),
suggesting the AGN hosts are dynamically similar to normal elliptical 
galaxies.
{\it Right:} Host galaxy magnitudes are similar for radio-loud AGN 
spanning
nearly 5 orders of magnitude in luminosity.
(Figures from O'Dowd et al.\ 2002.)
\vspace{-0.2cm}
}
\end{figure}

Inspired by this result, we investigated the relation 
of black hole mass to observable properties like AGN luminosity. 
There were claims in the literature that these were correlated, 
but these claims were either wildly optimistic or resulted from 
the use of the AGN luminosity to derive the black hole mass 
(thus the tautology that AGN luminosity is correlated with itself). 
Instead, when we collected several hundred black hole mass estimates, 
we found that there is essentially no correlation with AGN luminosity
(Woo \& Urry 2002a; Fig.~7). 
There is a rough trend such that the Eddington limit provides 
an approximate upper limit to the AGN luminosity, but apart from that, 
AGN appear at all masses and luminosities. 
At a minimum, this means that there is a wide range of 
Eddington ratios in AGN. The Eddington ratio does not correlate 
with mass or luminosity either. Such limits as are seen in the 
plots (Woo \& Urry 2002a) are selection effects, either from the 
survey flux limit or from the steepness of the luminosity function 
(high luminosities and/or high black hole masses are rare) or from the 
likely mis-identification of AGN as normal galaxies (when the nucleus 
reaches low enough luminosity relative to the host galaxy). 
Selection effects are particularly nasty; 
almost all the structure seen in the mass versus luminosity
plots for AGN derive from the original flux limits and/or volume limits
(see Woo \& Urry 2002a for details). 

\begin{figure}[t]
\label{fig:LMbh}
\plottwo{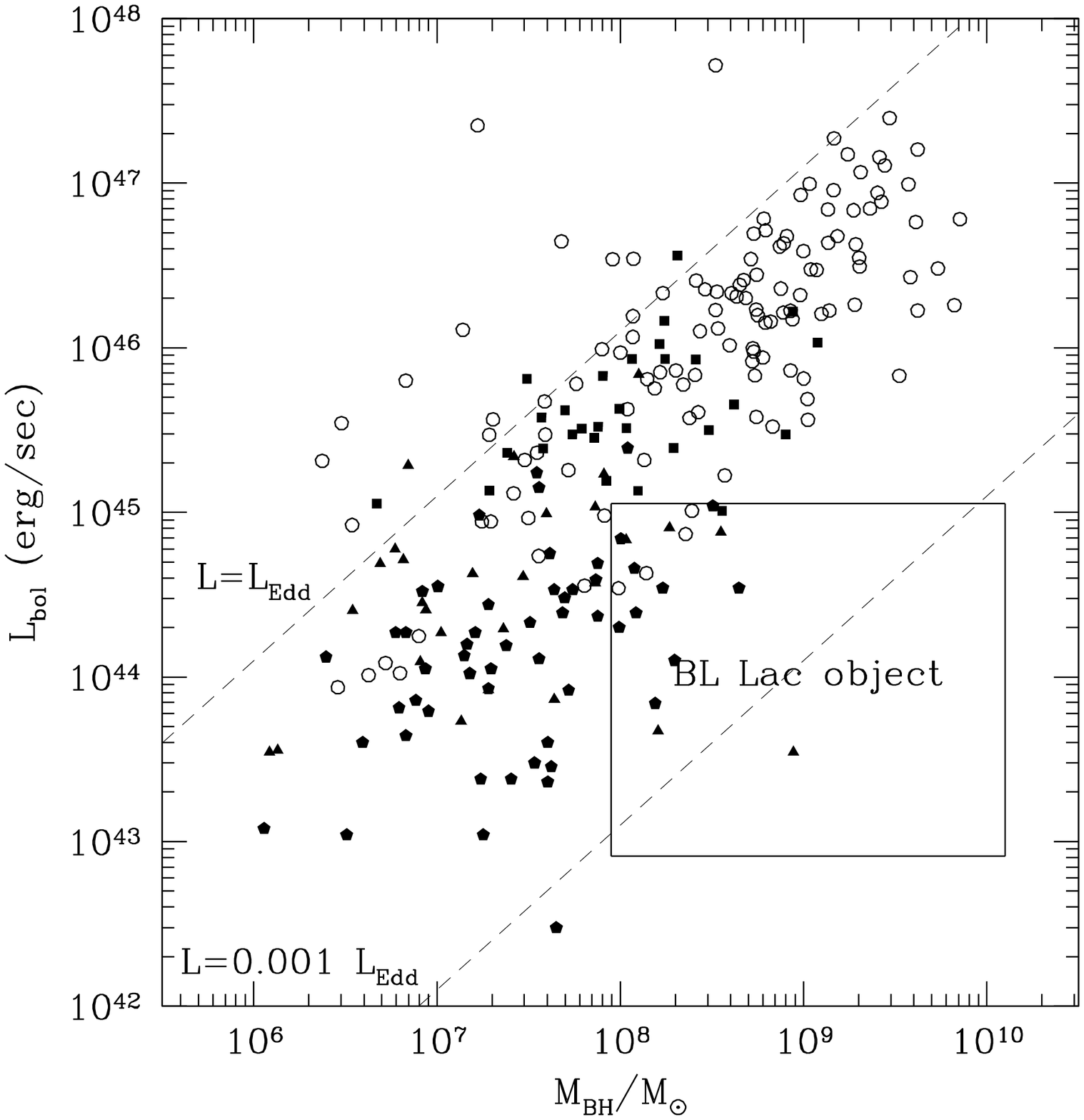}{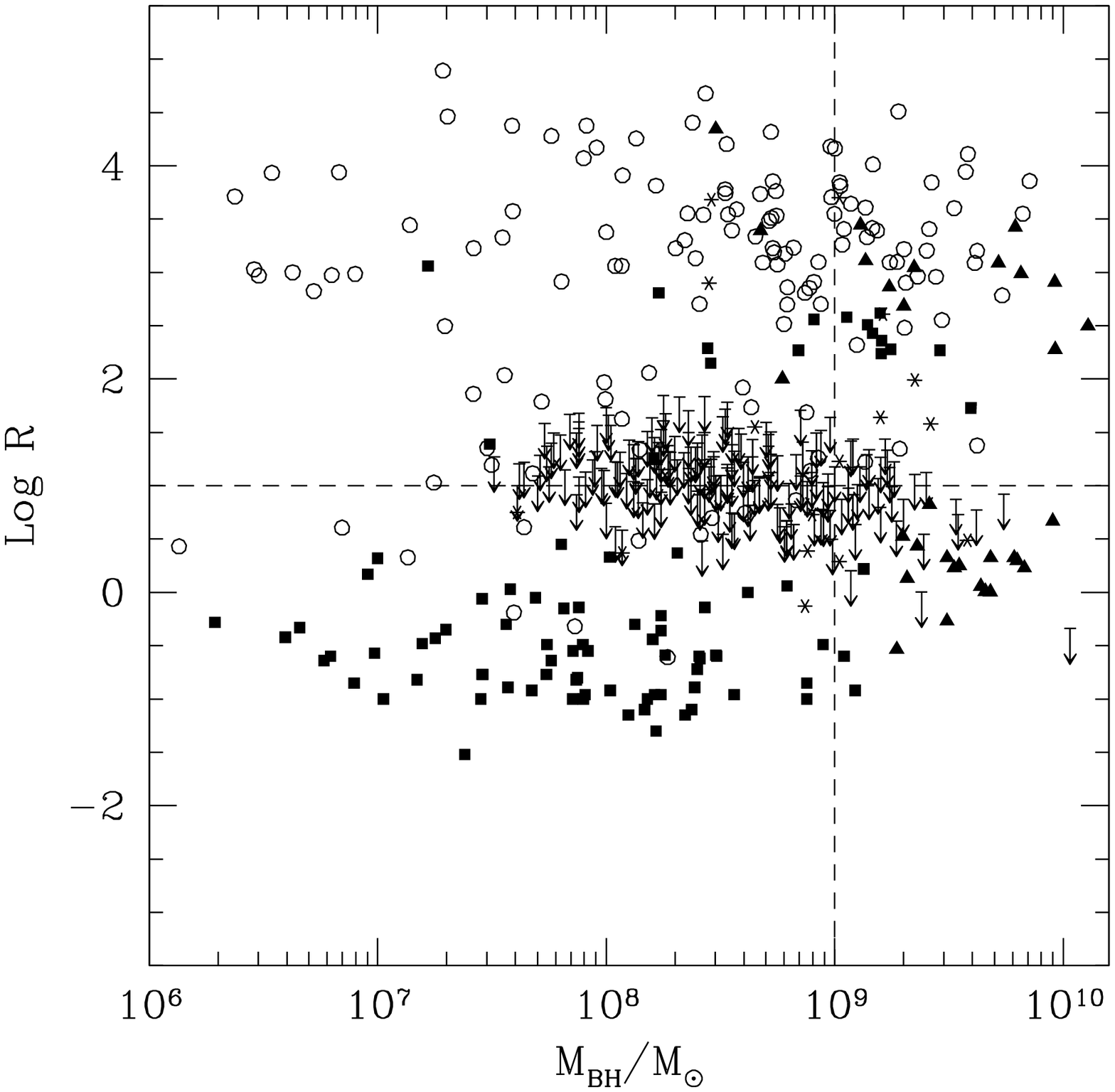}
\caption{
{\it Left:} 
Bolometric luminosity versus black hole mass for 234 AGN
(Woo \& Urry 2002a).
For a given black hole mass, there is a large range of bolometric
luminosities, spanning three or more orders of magnitude.
The Eddington limit defines an approximate upper limit
to the luminosity,
but the absence of objects from the lower right of the diagram
(low luminosity, high mass AGN) is a selection effect.
Symbols are {\it open circles:} radio-loud quasars; {\it filled squares:} radio-quiet
quasars; {\it filled triangles:} Seyfert 1; {\it filled pentagons:}
Seyfert 2.
{\it Right:} 
Radio loudness versus black hole mass for 452 AGN
shows no correlation between radio loudness and black hole mass
(Woo \& Urry 2002b).
In particular, at high mass ($M > 10^9 M\odot$) there
are similar distributions of black hole mass
for radio-loud ($R>10$) and radio-quiet AGN.
Symbols are {\it squares:} PG quasars, all at $z<0.5$;
{\it circles:} AGN at $0<z<1$;
{\it triangles:} high-redshift quasars ($2<z<2.5$) from McIntosh et al.\ (1999);
{\it stars:} LBQS quasars, $0.5<z<1$;
{\it arrows:} upper limits for LBQS quasars, $0.5<z<1$.
\vspace{-0.3cm}
}
\end{figure}

We also investigated whether radio power could be related to 
black hole mass (Woo \& Urry 2002b), as has been suggested by
a number of authors (Franceschini, Vercellone, \& Fabian 1998; 
McLure et al.\ 1999; Lacy et al.\ 2001; Nagar et al. 2002; Jarvis \& McLure 2002). 
However, there are radio-loud AGN with low black hole masses
(Ho 2002; Oshlack, Webster, \& Whiting 2001),
and we found radio-quiet AGN with very high black hole masses, 
in the range $10^9-10^{10} M_\odot$. Thus there appears to be 
no requirement for radio-loud objects to have high black hole masses, 
nor does high black hole mass imply radio loudness. 

So, contrary to some previously published claims:\vspace{0.1cm}\\
$\bullet$~There is no correlation of AGN luminosity or Eddington ratio with black hole mass, apart from an upper envelope around $L/L_{Edd} \sim 10-100$.\vspace{0.1cm}\\
$\bullet$~There is no correlation of Eddington ratio with AGN luminosity. 
It is often asserted that the most luminous AGN (quasars) have high 
Eddington ratios while low-luminosity AGN have low Eddington ratios. 
There is a gross trend in this direction, but in our samples it can be ascribed 
entirely to selection effects (cf., Merloni et al. 2004, this volume).\vspace{0.1cm}\\
$\bullet$~There is no relation between radio loudness and black hole mass. 

\section{Summary}
\label{sec:summary}
\vspace{-0.05cm}
Decades of research have documented the inherently anisotropic
nature of AGN, and thus unification must be the case to at
least some extent. It remains to be seen how much intrinsic variation there 
is in the fundamental AGN properties (black hole mass, accretion rate, 
and black hole spin) and in AGN structure.
It is likely that obscuration and/or AGN structure evolves, and that 
unification locally is different in detail from
unification at the quasar epoch. 

Radio unification and the effects of relativistic beaming are well 
understood and seem to explain well the family of radio-loud AGN. 
Lorentz factors of order 10 explain well the numbers of steep-spectrum 
and flat-spectrum radio-loud quasars relative to radio galaxies, 
and also fit with the observed superluminal velocities and 
the core dominance parameters. 

The number of obscured AGN is unclear. There should be a substantial 
population of them at $z\sim1$, as suggested from the X-ray background. 
Deep multiwavelength surveys sensitive to obscured AGN are just 
now being carried out, and these questions will surely be answered 
in the next year. These surveys will also shed light on the 
relation of obscured AGN to starbursts and ultra-luminous IR 
galaxies, which Sanders et al.\ (1988) long ago 
suggested were shrouded quasars in formation. The links found since 
then between the host galaxies and the active nucleus lend support 
to this suggestion, and direct observation of the link is within reach.

Finally, black hole mass is not an important parameter as far as 
defining the appearance of an AGN. It is essentially uncorrelated 
with AGN luminosity, contrary to popular myth. Nor is it correlated 
with radio power --- whatever the jet formation mechanism, 
it is not strongly dependent on black hole mass. 
The distribution of jet power is one of the remaining great unknowns. 
Studies of blazar jets and blazar demographics will answer this question.

We are now poised to take unification to the next level, and 
to have a Grand Unification of AGN with galaxies. Within the next 
few years, we should be able to determine whether AGN are indeed a 
common phase in galaxy evolution, and to probe the co-evolution
of galaxies and black holes. 

\acknowledgements 
The work of my graduate students figures prominently in this paper, 
particular Matthew O'Dowd (U.\ Melbourne), and my Yale students, 
Jong-Hak Woo, Ezequiel Treister, Brooke Simmons, and Jeff Van Duyne. 
I also thank my postdoc, Eleni Chatzichristou, who has worked hard 
on the GOODS project, as well as the entire GOODS team, particularly 
the AGN subgroup, and I thank my blazar collaborators. 
This work was made possible by support from grants
HST-GO-09425.13-A and NAG5-12873.

\end{document}